# Exploring fair access to quantum computing

Benedict Lane and Pieter Vermaas

Philosophy and Ethics of Technology,

TU Delft

The Netherlands

j.b.h.lane@tudelft.nl; p.e.vermaas@tudelft.nl

Abstract:

This short article – commissioned as part of the Quantum Flagship project OpenSuperQPlus – introduces the topic of fair access to quantum computing time and argues for its relevance in current European technology policy debates. Choices about access we make today affect which applications of quantum computing are developed in the future, and by whom. Among the issues at stake are: Europe's technological autonomy and sovereignty; Europe's future economic competitiveness; knowledge security and the future of open science; and delivering upon the EU's founding values and aims. The results of a survey of OpenSuperQPlus researchers reveal prevalent contractualist and transactional framings of access to European quantum computing capacity among those developing the technology. The article concludes with a call to go beyond framing access to quantum computing as a zero-sum-game, and to explore the possibility of an access regime premised on openness and solidarity.[1]

[1] This paper is a merger of three texts published at the website of the OpenSuperQPlus project: part 1 appeared at https://opensuperqplus.eu/news/exploring-fair-access-to-quantum-computing on October 14, 2025; part 2 appeared at https://opensuperqplus.eu/news/exploring-fair-access-to-quantum-computing-part-2 on October 24, 2025; and part 3 appeared at https://opensuperqplus.eu/news/exploring-fair-access-to-quantum-computing-part-3 on October 30, 2025.

# 1. Equity and policy implications

## Why care about access to quantum computing?

OpenSuperQPlus – the European effort to build a large quantum computer with superconducting qubits – is making rapid progress and will soon offer larger functioning prototypes. With this development, questions of who gets access to quantum computing are becoming urgent. Choices made today about access to quantum computing will shape the development, impact, and public acceptance of quantum computing technologies into the future. Striving to get it right from the start - and remaining open to necessary improvements in the future - is essential for ensuring these potentially powerful technologies serve the public good.

## Meaningful use

Choices about access directly influence which applications of quantum computing are explored and exploited, and by whom. With the potential for both positive and negative impacts – from innovative drug discovery to invasive digital surveillance – controlling access becomes a way of steering use. Equitable access can help prioritise scientifically, industrially, and socially valuable use cases and guard against harmful ones.

Choices about access determine which use cases are identified in the first place. Naïve access policies can concentrate use cases within narrow, well-resourced sectors, risking what the World Economic Forum (2024) has called a "two-speed" development of quantum applications – whereby applications of particular interest to certain dominant groups are developed quicker than applications that serve the interests of a more inclusive range of stakeholders. For example, pharmaceutical companies invest heavily in quantum computing to speed up drug discovery and drive profits, while socially beneficial applications of quantum computing with looser links to industry, such as climate modelling and water stress forecasting, languish from underinvestment. Broader access enables identification of overlooked problems and novel solutions. Hackathons and open competitions like Pasqal's [re]Generative Quantum Challenge (2023) show how grassroots participation can uncover socially beneficial uses.

## Awareness, talent, and ecosystem development

Access is not just about short-term use – it is about cultivating long-term social and technical capacities. An access framework that includes non-experts, for example, can help build quantum literacy across society, starting with students and young people. Initiatives like the Open Quantum Institute's Education Consortium show how early access fosters understanding and participation. Smart access decisions can also upskill existing experts, reduce talent bottlenecks, and prevent talent and knowledge inequalities from deepening as quantum computing technologies advance – this is essential for developing a robust European quantum ecosystem.

Major European funding initiatives, such as the OpenSuperQPlus project, demonstrate the strong research-policy relevance of further developing quantum computing. However, this future technology must be built on a socially inclusive and secure political foundation to fully unfold its innovation potential across scientific, societal, and industrial domains.

# 2. Balancing open science, security, and prosperity in access to quantum technology

## Open science, individual autonomy, and democratic legitimacy

Open access underpins open science, expanding opportunities for productive collaboration and encouraging the dissemination of useful findings. Public access promotes individual autonomy – empowering people to learn, experiment, and innovate, reducing barriers to social mobility. The perception of fair access also supports democratic legitimacy: when access is seen as fair, public trust in quantum technologies grows. The scientific progress of OpenSuperQPlus builds on this approach to equitable access. By using interoperable access models, it contributes to the translation of technologies and fosters research that is not dependent on the investment of individual players.

## Security, safety, and sovereignty

However, responsible access also requires careful attention to security, safety, and sovereignty concerns. In their recent report on export controls and quantum technologies, the Quantum Strategy Institute (2025) identifies several key reasons to restrict access to quantum computing on security and strategic grounds. Uncritical allocations of access to quantum computing may (i) put the power to decrypt strategically sensitive communications into the wrong hands and (ii) forego strategic technological advantages over competitors, for example, through reverse engineering. In their report on the potential impact of quantum technologies on law enforcement, Europol (2024) highlights the possibility of quantum computers efficiently guessing user passwords to combat abuse and terrorism offences, which have a major digital component. However, in the wrong hands, the capacity to access password-protected data is clearly problematic. EU member states are also under a legal obligation to limit exports of dual-use technologies falling under the Wassenaar Arrangement (2019). Striking the right balance between openness and security, within the framework of the law, is an important task with scientific, economic, and societal implications.

## A quantum computer for Europe

The OpenSuperQPlus quantum computer will be a tangible symbol of Europe's ambition to develop technological sovereignty and strategic autonomy. The Draghi Report (2024) – the European Commission's report on the future of EU competitiveness – singled out cloud access to quantum computing provided from abroad (particular from US tech giants) as a critical European vulnerability, and domestic development of quantum computing capacity as crucial to Europe's future scientific and industrial competitiveness on the global stage. The OpenSuperQPlus computer can therefore be seen as an important strategic infrastructure at the heart of Europe's technological and economic ambitions. Equitable access would ensure that all member states and citizens – with access granted based on need rather than on a transactional basis – can contribute to and benefit from the development of quantum computing technologies in Europe. This will strengthen social cohesion and solidarity across the Union.

At the national level, needs-based allocation of access to the OpenSuperQPlus computer can support national talent development and research initiatives, providing tailwinds for national quantum strategies (OECD, 2025) – Austria, the Czech Republic, Denmark, Finland, France, Germany, Hungary, Ireland, Italy, Luxembourg, the Netherlands, Spain, and Sweden have all implemented or drafted national quantum strategies as of 2025.

## European values for the quantum future

The OpenSuperQPlus computer also represents an opportunity to promote core European values. The EU's founding charter identifies democracy, human rights, and equality as foundational; the EU's digital principles, as laid out in the European Declaration on Digital Rights and Principles (2023), encompass inclusion, participation, and the empowerment of individuals through access to digital infrastructure. The EU's Quantum Manifesto (2016) seek to advance European quantum for the benefit of society.

Allocating access to the OpenSuperQPlus computer in a way that is consistent with these values can contribute to upholding them in meaningful ways. Inclusive access policies can ensure that the computer is employed in service of diverse societal needs, prevent monopolisation, and give smaller institutions, underserved regions, and marginalised groups more sway over Europe's quantum future. As observed, allocating quantum computing time to non-expert educational users or mission-driven projects can lead to sustainable solutions to pressing societal problems. Likewise, openness and transparency in the allocation process empowers EU citizens to scrutinise the uses to which the OpenSuperQPlus computer is put, ensuring democratic accountability and public trust.

### Scaling up Europe's quantum technology

Europe has long excelled in fundamental research on quantum, from the paradigm-shifting breakthroughs of the 20th century to Europe's current array of quantum tech start-ups (many of which – Alice & Bob, IQM, OrangeQS, Qruise, QuantWare – are partners in OpenSuperQPlus). By making choices about access that further promote European scientific excellence, we can consolidate and strengthen Europe's world-leading position at the cutting edge of fundamental and applied quantum research.

However, the Draghi Report notes a persistent weakness in scaling up European start-ups into world-leading technology companies. Intelligent choices about access to the OpenSuperQPlus computer can contribute to addressing this weakness in the context of quantum computing. By prioritising ambitious projects aiming at more advanced levels of technological- and market-readiness, the OpenSuperQPlus computer can contribute to fostering much-needed industrial scaling-up in the European quantum technology sector. An ambitious-yet-principled access framework would ensure that this scaling up occurs in a balanced and mutually beneficial way across the EU – promoting growth across borders, fostering a truly European quantum industry.

## 3. Insights from OpenSuperQPlus experts

### What do OpenSuperQPlus researchers think?

In a recent survey, we invited OpenSuperQPlus researchers to share their views regarding access to the quantum computers OpenSuperQPlus is developing. This was also the subject of a lively panel discussion during our progress meeting in Helsinki, May 2025. We inquired about researchers' views along three dimensions: (i) prioritising users and uses, (ii) distributing benefits, and (iii) decision-making.

### Prioritising users and uses

First, we asked OpenSuperQPlus researchers for their views about who should get access to the computer and for what purpose. Perhaps unsurprisingly, the results reveal a strong preference for research and education-oriented uses. Over 60% ranked quantum computing R&D or fundamental quantum physics as their top use-priorities, and 90% regarded the potential to contribute to scientific knowledge as an important or extremely important factor in determining access. In contrast, only 8% recognised a legitimate claim to access for the public.

Justifications centred both on scarcity of quantum computing as a resource and the technical burden of broader access. However, a minority saw long-term value in broad public access. One respondent cited IBM's model of free and early access as key to cultivating "a vibrant community" of users, translating "into greater interest and broader engagement” in quantum computing technology. Indeed, during the panel in Helsinki, it was pointed out that dedicated communities of amateur users sprung up around early digital computers in the 1990s, and that these communities played an integral role in propelling the technology to public prominence. Perhaps, if sufficient access were available, quantum computers could play a comparable community-oriented role in the future? Another respondent saw benefits of open access for improving the technology itself, suggesting that the OpenSuperQPlus computer “could be an openly accessible platform for users who want to share their results, and for the produced data to feed back into the improvement of the system.” In these ways, while privileged access would directly benefit scientists, more open access might still benefit scientists indirectly, by cultivating interest in and demand for quantum computing as well as providing a valuable source of data for improving technical performance.

## Distributing benefits

Second, we asked researchers about who should benefit from quantum computing, and on what basis. The uniformity of responses was striking. While over 95% of respondents regarded the potential of the OpenSuperQPlus computer to promote public goods within the EU as either important or extremely important for determining access, support dropped sharply for promoting goods beyond Europe. More than 50% of respondents saw little relevance in the potential of the OpenSuperQPlus computer to advance EU interests abroad or to address global inequality. 50% of users viewed the potential for non-EU users to make profitable use of the computer as completely irrelevant. In contrast, the same percentage of respondents viewed the potential for EU-based users to make profitable use of the computer as either important or extremely important.

Justifications offered by respondents for such a marked difference in priority between EU and non-EU beneficiaries reflected a quasi-contractual view: the computer was funded by EU taxpayers, and it should therefore primarily benefit them – this applied equally to institutions promoting the public good or corporations making a profit. Such a line of thought is based on an implicit assumption: that quantum computing is a zero-sum game – in other words, that if others benefit from our quantum computer, we miss out. However, in the words of one respondent, it is important to remember that “we also profit from helping others for the common good.” Perhaps these results and reflections might serve as food for thought regarding the fairest global distribution of the benefits of quantum technologies developed in Europe.

## Decision-making

Lastly, we asked researchers to reflect on the values and principles that ought to underlie how decisions about access to the OpenSuperQPlus computer are made. Over 70% of researchers agreed that transparency was the most important factor in decision-making about access. Again, responses reflected a quasi-contractual view, consistent with researchers' views about distributing benefits: as one respondent put it, "since it is a public funded infrastructure, the public should know how it is being used and for what purposes." Beyond this, there was significant diversity among respondents about which principles and values should underlie how choices about access are made: scientific merit, fair competition, equality of opportunity, economic growth, and social benefit were all invoked as decision-making criteria. However, there was a common recognition that OpenSuperQPlus occupied a unique position among quantum computing platforms: in the words of one respondent, "We have an opportunity to make OSQPlus different than just a competitor to commercial systems (a competition we wouldn't win, by the way)."

### Advancing Technology with Social Responsibility

Creating a quantum computer with superconducting qubits is a job for specialists. Making decisions about allocating the computing time made available by such computers is something we should do with all stakeholders involved. OpenSuperQPlus does both. It will deliver prototype quantum computers and initiate and develop governance on the allocation of quantum computing time. We are reaching out to stakeholders to collect and aggregate the social and moral values they put forward for this allocation. We did so already for a special stakeholder group – the researchers making available the OpenSuperQPlus quantum computers. And we will reach out to industry, research, society , and governance, devoting equal care to arrive at a societally meaningful development of quantum computing in Europe through principled allocation of access.